\documentclass[onecolumn]{aastex631}
\usepackage{mathtools}
\usepackage{amsmath}
\usepackage{soul}
\usepackage{lipsum}
\usepackage{gensymb}
\usepackage{rotating}
\usepackage{float}

\newcommand\kms{$\mathrm{km\ s^{-1}}$}
\newcommand\Rsun{$\mathrm{R_{\odot}}$}

\newcommand\oxy{\mathrm{O^{7+}/O^{6+}}}

\newcommand\Sp{\mathrm{S_p}}

\newcommand\sigmac{$\mathrm{\sigma_C}$}

\newcommand\yeff{\gamma_{eff}}
\newcommand\alf{Alfv\'en}
\newcommand\alfic{Alfv\'enic}
\newcommand\alfty{Alfv\'enicity}

\newcommand\elsasser{Els\"{a}sser}

\shorttitle{Proton specific entropy as a proxy for {$\oxy$}}

\begin{document}

\title{Proton specific entropy as a proxy for the {$\oxy$} charge state ratio over heliocentric distance}

\correspondingauthor{Jack D. Collard}
\email{jdcollard@berkeley.edu}

\author[0009-0005-6599-3255]{Jack D. Collard}
\affiliation{Department of Physics, University of California, Berkeley, Berkeley, CA 94720-7300, USA}
\affiliation{Space Sciences Laboratory, University of California, Berkeley, CA 94720-7450, USA}

\author[0000-0002-8475-8606]{Tamar Ervin}
\affiliation{Department of Physics, University of California, Berkeley, Berkeley, CA 94720-7300, USA}
\affiliation{Space Sciences Laboratory, University of California, Berkeley, CA 94720-7450, USA}

\author[0000-0003-4437-0698]{Ryan M. Dewey} 
\affiliation{Department of Climate and Space Sciences and Engineering, University of Michigan, Ann Arbor, MI 48109, USA}

\author[0000-0002-8748-2123]{Yeimy J. Rivera} 
\affiliation{Center for Astrophysics $\vert$ Harvard \& Smithsonian, 60 Garden Street, Cambridge, MA 02138, USA}

\author[0009-0002-0149-8747]{Aidan J. Nakhleh} 
\affiliation{Department of Climate and Space Sciences and Engineering, University of Michigan, Ann Arbor, MI 48109, USA}

\author[0000-0002-1628-0276]{Jean-Baptiste Dakeyo}
\affiliation{Space Sciences Laboratory, University of California, Berkeley, CA 94720-7450, USA}

\author[0000-0002-6145-436X]{Samuel T. Badman}
\affiliation{Center for Astrophysics $\vert$ Harvard \& Smithsonian, 60 Garden Street, Cambridge, MA 02138, USA}

\author[0000-0002-4625-3332]{Trevor A. Bowen}
\affiliation{Space Sciences Laboratory, University of California, Berkeley, CA 94720-7450, USA}

\author[0000-0002-0675-7907]{John W. Bonnell}
\affiliation{Space Sciences Laboratory, University of California, Berkeley, CA 94720-7450, USA}

\author[0000-0003-1692-1704]{Nicholeen M. Viall} 
\affiliation{NASA Goddard Space Flight Center, Greenbelt, MD, USA}

\author[0000-0003-1611-227X]{Susan T. Lepri} 
\affiliation{Department of Climate and Space Sciences and Engineering, University of Michigan, Ann Arbor, MI 48109, USA}

\author[0000-0001-5956-9523]{Jim M. Raines} 
\affiliation{Department of Climate and Space Sciences and Engineering, University of Michigan, Ann Arbor, MI 48109, USA}

\author[0000-0002-1989-3596]{Stuart D. Bale}
\affiliation{Department of Physics, University of California, Berkeley, Berkeley, CA 94720-7300, USA}
\affiliation{Space Sciences Laboratory, University of California, Berkeley, CA 94720-7450, USA}

\begin{abstract} 

While the fast solar wind has well-established origins in coronal holes, the source of the slow solar wind remains uncertain. Compositional metrics, such as heavy ion charge state ratios are set in the lower corona, providing insights into solar wind source regions. However, prior to the launch of Solar Orbiter, in situ measurements of heavy ion charge state were limited to distances of 1 AU and beyond. We investigate proton specific entropy as a proxy for the oxygen charge state ratio ({$\oxy$}),which generally becomes frozen-in below $\sim$1.8~{\Rsun}, leveraging observations from Solar Orbiter's Heavy Ion Sensor and Proton and Alphas Sensor covering 0.28 to 1 AU. Our analysis confirms a strong anti-correlation between specific entropy and the oxygen charge state ratio that persists over a broad range of distances in the inner heliosphere. We categorize observed solar wind into fast solar wind, slow {\alfic} solar wind, and slow solar wind, identifying clear distinctions in specific entropy values and charge state ratios across these types. The work demonstrates the potential to use proton specific entropy as a classifier of solar wind source regions throughout the heliosphere. By establishing the $\Sp$-$\oxy$ relationship and quantifying its radial dependence, the specific entropy can be used as a quantity to identify the solar wind source region in the absence of in-situ charge state measurements. This motivates future studies as to the applicability of this proxy to near-Sun observations (such as Parker Solar Probe) and throughout the inner heliosphere.

\end{abstract}

\section{Introduction} \label{sec: intro}

The solar wind is comprised of streams of charged particles, composed of protons, electrons, alpha particles, and traces of heavy ions, that are continuously escaping from the corona \citep{Parker-1958} and flowing supersonically through the heliosphere. While the fast solar wind (FSW; speeds above 500 {\kms} at 1 AU), has been shown to emerge from coronal holes \citep[CHs;][]{McComas-1998, McComas-2008}, it is still uncertain as to the contribution of different structures to the slow solar wind \citep[][]{Viall-2020}. Identifying the source of the slow solar wind is a complex problem, due to various difficulties with combining models, observations and theory. Compositional metrics, such as heavy ion charge state ratios and elemental abundances, have been found to be a potential tracer of the solar wind’s source region(s) which could help distinguish the different sources contributing to the slow wind \citep[e.g.][]{vonSteiger-2000, Rivera-2022wp, Ervin-2024CH, Rivera-2024SB}. Heavy ion charge states measured in the solar wind at large heliocentric distances are related to their solar sources (coronal hole, quiet Sun, active region) through their relative densities which can reflect the thermal structure of their coronal origins. While the elemental composition measured in the heliosphere (specifically the degree to which the first ionization potential bias is observed) is broadly governed by the magnetic topology and wave activity of the solar wind's coronal source.

In the solar wind, ion charge state ratios remain constant after some radial distance from the Sun, due to the so-called \lq{}freeze-in\rq{} process \citep{HundhAUsen-1968, Owens-2018}. As the solar wind moves outward from the Sun, the electron density and temperature decrease until it reaches a point, the \lq{}freeze-in\rq{} point, where ionization and recombination can no longer occur and the ion abundance ratios remain fixed after this distance \citep{Owocki-1983}. The freeze-in point is low in the corona, typically $<1.8$~{\Rsun}, as shown by observations and modeling \citep{Landi-2012b, Boe-2018, Gilly-2020, Szente-2022, Riley-2025}. This process allows the use of these quantities to determine information about the coronal plasma from where the solar wind emerges \citep{Buergi-1986, Chen-2003}. The variability in the timescale associated with ionization and recombination processes is a consequence of changing solar wind conditions up to their freeze-in altitude as well as among solar sources that ultimately leads to differences in heavy ion ratios measured in-situ. These ratios contain information about the thermodynamic history of the plasma parcel specific to its source which can be used as tracers of lower coronal origin.

It has been established that fast solar wind (FSW) streams, originating in CHs, show ion composition ratios (e.g. $\mathrm{C^{6+}/C^{5+}, C^{6+}/C^{4+}, O^{7+}/O^{6+}}$ etc.) that are lower, consistent with quick acceleration of plasma from cooler, lower density coronal hole structures \citep[][]{Feldman-1998, Doschek-1998, vonSteiger-2011, Ervin-2024CH}. In comparison, the slow solar wind (SSW) often has largely varying charge state ratios that have been related to varied coronal source regions \citep{Raymond-2001, Brooks-2015, Liewer-2003, Culhane-2014}. Total elemental composition (e.g. the amount of Fe or O or C) is also dependent upon the solar wind source structure. Solar wind parcels that emerged quickly (e.g. on open field lines) have elemental abundances that are similar to photospheric levels, while wind that spent time in topologically closed regions show enhancements in low-FIP (First Ionization Potential) elements. The processes that lead to the fractionation of the plasma, and thus differences in elemental abundances, are also thought to take place in the chromosphere, where the plasma is only partially ionized, and lower corona \citep[e.g.][]{Laming-2015}, however FIP bias measurements will not be the focus of this work.

While composition has been shown to be an excellent identified of source regions, transport processes and stream interactions that develop between the low corona and where it has been historically measured (near 1 AU) can make interpretations difficult. Ideally, one would want to sample the charges states closer to their solar sources where they can be more accurately connected to the places they were formed, thus eliminating much of its evolution as a factor in disentangling source region signatures. In particular, Parker Solar Probe \citep[PSP;][]{Fox-2016}, while reaching distances very close to the Sun, lacks measurements of heavy ion charge states (such as {$\oxy$}) in the ambient wind, a critical piece to building a full picture of the lower coronal source regions of the solar wind, as the composition and plasma parameters of the SSW are generally more variable compared to the FSW, making it more difficult to distinguish and trace the SSW to its (variable) source(s). These compositional measurements are also difficult to make, due to challenges with spacecraft mass budgets (among other factors) and there are limited observations in the near-Sun environment. It is therefore of interest to find proxies for the oxygen charge state ratio ({$\oxy$}) with application for source region identification.

\citet{Pagel-2004, Xu-2015} showed that proton specific entropy, $\Sp=T_p/N_p^{(\gamma - 1)}$ (where $T_p$ is the proton temperature and $N_p$ is the proton density), is strongly anti-correlated with {$\oxy$}, despite not being conserved through radial evolution. However, this correlation has never been established closer to the Sun. Variability in proton specific entropy is thought to be generated in the lower corona due to its strong correlation with frozen-in properties (like the {$\oxy$} ratio). Through a cross-correlation analysis of 1 AU measurements of FIP bias (e.g. $\mathrm{Fe/O}$) and heavy ion charge states with proton specific entropy, \citet{Nakhleh-2025} showed that the correlation between entropy and charge state to be much stronger than the entropy-FIP bias correlation. As the charge state ratios and the FIP bias are caused by different physical processes (and tied to magnetic topology, not just energization in the low corona), \citet{Nakhleh-2025} showed that the entropy variability associated with the transition from the partly collisional chromosphere to the collisionless corona, and is frozen-in between 1.4 and 1.8~{\Rsun}. Additionally, if one can determine the ``correct" polytropic index ($\gamma$) to describe the evolution of the temperature and density profile of a solar wind stream, $\Sp$ is expected to be conserved through radial evolution, making it a good potential proxy to study.

Motivated by the relationship shown in \citet{Pagel-2004}, and the results of entropy freeze-in \citep{Nakhleh-2025}, we examine the relationship between specific entropy and the oxygen charge state ratio using Solar Orbiter \citep{Muller-2020} heavy ion measurements from the Heavy Ion Sensor \citep[HIS;][]{Livi-2023} to understand the correlation and potential use of {$\Sp$} as a source region classifier. Solar Orbiter provides the unique ability to study this correlation as a function of radial distance and with multi-point measurements (in combination with spacecraft sans heavy ion instruments) during spacecraft conjunctions in future studies.

In Section~\ref{sec: correlation} we describe the relation between specific proton entropy and the oxygen charge state ratio using Solar Orbiter observations. We discuss the dependence on the choice of polytropic index value ($\gamma$) and the the relation with cross helicity. In Section~\ref{sec:radial}, we discuss the dependence on radial distance, and propose an effective polytropic index ($\yeff$) that gives a ``conserved" $\Sp$ over radial distance. In Section~\ref{sec: categorization} we discuss the relationship between specific entropy, oxygen charge state ratios with the wind type and associated source region. Lastly, in Section~\ref{sec: conclusions} we outline our main takeaways and argue for the need for additional inner heliospheric charge state and elemental composition observations to further constrain solar wind sources and acceleration mechanisms. 


\section{Correlation between proton specific entropy and {$\oxy$}} \label{sec: correlation}

We use observations from Solar Orbiter from January 2022 through April 2023 to study the relationship between the specific proton entropy ($\Sp=T_p/N_p^{(\gamma - 1)}$) and oxygen charge state ratio ({$\oxy$}) closer to the Sun than ever before. Magnetic field measurements and plasma moments (density, velocity, and temperature) on Solar Orbiter come from the Magnetometer \citep[MAG:][]{Horbury-2020} and the Proton and Alphas Sensor (PAS) aboard the Solar Wind Analyzer \citep[SWA;][]{Owen-2020} respectively. HIS provides 10-minute integrated measurements of heavy ion charge state ratios ({$\oxy$} is used in this study, but many others as well) as well as uncertainties that are used in this analysis. The time period is chosen to reflect the period where HIS makes continuous measurements that have been calibrated and validated by the instrument team. Appendix~\ref{app:data-cleaning} describes the process through which we produced a combined dataset between these instruments.

We examine the relationship between specific entropy and the {$\oxy$} ratio to understand how the specific entropy can be used as a source indicator. There are already well-established categories for the solar wind that allow us to relate in-situ parameters to their source. Typically, the solar wind is classified as fast ($v > 500$ {\kms} at 1 AU) and slow ($v < 500${\kms} at 1 AU) with perhaps further sub-categorization by other parameters such as the level of {\alfty} (or cross helicity). The slow wind has multiple possible sources and thus requires multiple in-situ parameters to establish a given source \citep{Viall-2020}.

\begin{figure}[ht!]
    \centering
    \includegraphics[width=\columnwidth]{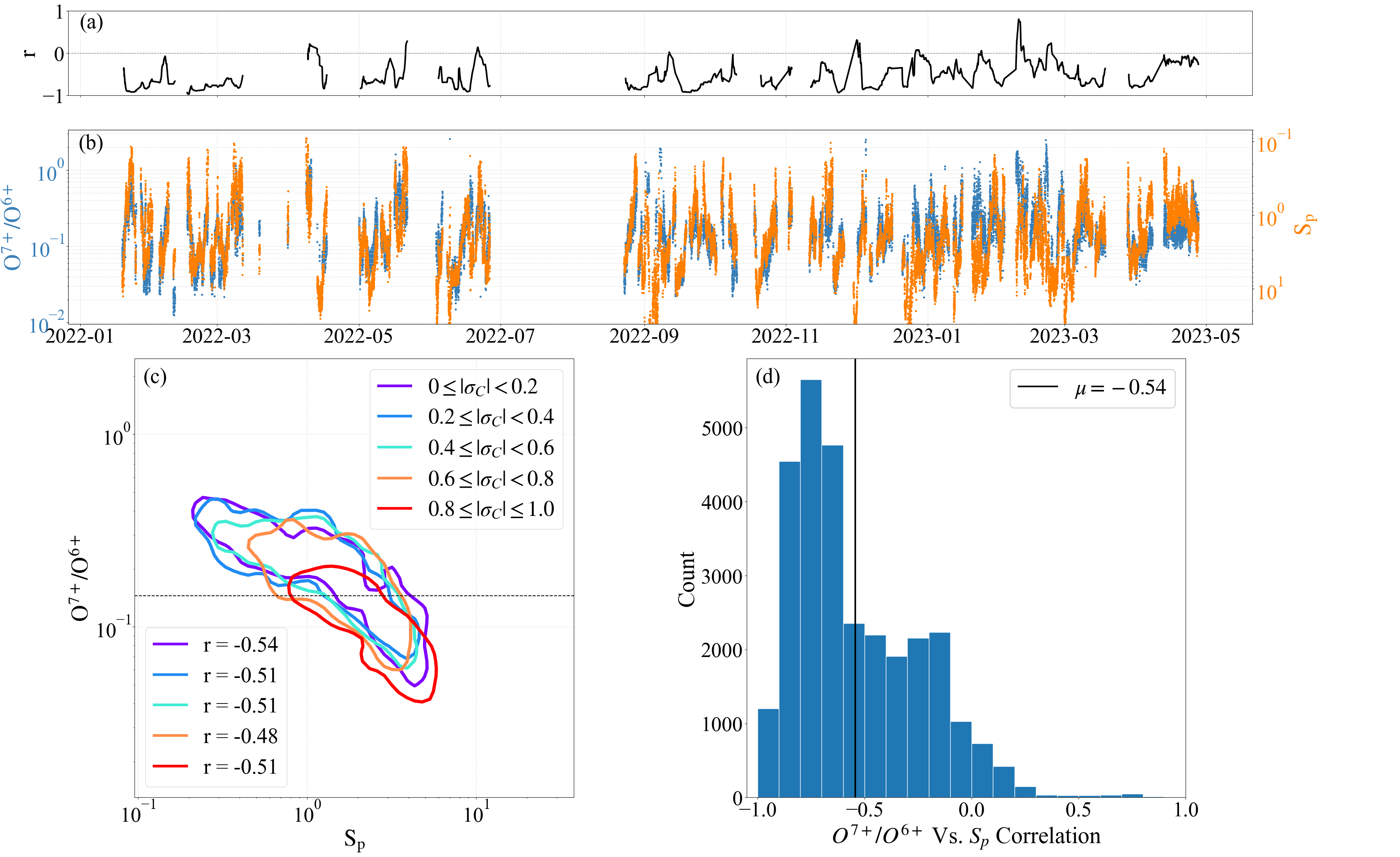}
    \caption{Charge state ratio temporal evolution and correlation with proton-specific entropy as measured by Solar Orbiter from January 2022 through April 2023. Panel (a): Rolling Pearson correlation coefficient (r) between log({$\Sp$}) and log({$\oxy$}), computed using a 7-day time-based window. Panel (b) Timeseries of {$\oxy$} (blue, left axis) and {$\Sp$} (orange, right axis) The {$\Sp$} axes has been flipped. Panel (c) Log-log distribution of {$\oxy$} versus {$\Sp$} separated into five cross-helicity intervals. Each colored contour encloses the densest 45\% of data within its bin, and corresponding log-log correlation coefficients are listed in the legend. The dashed horizontal line at {$\oxy$} = 0.145 denotes the upper threshold for solar wind emerging from CHs \citep{Zhao-2009}. Panel (d): Histogram of the rolling Pearson correlation values from panel (a). The vertical line marks the mean ($\mu \approx -.54$).}
    \label{fig: timeseries_correlation}
\end{figure}

Figure~\ref{fig: timeseries_correlation} presents the temporal and statistical relationship between {$\Sp$} and {$\oxy$} from January 2022 through April 2023. Panel (a) shows the rolling Pearson correlation coefficient (r) between log({$\Sp$}) and log({$\oxy$}), over a 7-day window, demonstrating the temporal consistency of their anti-correlation. Panel (b) displays the corresponding time series of {$\oxy$} compared with {$\Sp$}, where the anti-correlation of the two parameters is qualitatively evident. Panel (c) demonstrates a strong relationship between the parameters through a joint distribution of probability density contours, supporting the investigation of specific proton entropy as a proxy for the {$\oxy$} charge state ratio over varied cross helicity values. Panel (d) shows the distribution of the the correlation coefficients over time from panel (a), with most values clustering at $r \leq -0.5$. This distribution shows that the anti-correlation between the parameters in maintained across the majority of the time span.

Panel (c) is contours are colored by the absolute value of cross helicity, {\sigmac}, which indicates the level of imbalance ({\sigmac} $\neq$ 0) in the turbulence and if the wind is {\alfic}. {\alfic} fluctuations are ubiquitous in the FSW, and work has found the SSW to also be highly {\alfic} at times \citep[][and others]{DAmicis-2015, Ervin-2024CH, Ervin-2024SASW, Rivera-2025SSW}, known as slow {\alfic} solar wind (SASW). 

Cross helicity is calculated following the methods outlined in \citet{Ervin-2024SA, Ervin-2024SASW} and in Equation~\ref{eqn: sigmac} where $E^\pm$ is the energy associated with the {\elsasser} variables defined as: $\mathbf{z}^\pm = \delta \mathbf{v} \mp \mathrm{sign} \langle B_{r} \rangle \delta \mathbf{b}$, such that $\delta \mathbf{v} = \mathbf{v} - \langle \mathbf{v} \rangle$ and $\delta \mathbf{b} = \mathbf{b} - \langle \mathbf{b} \rangle$. $\mathbf{b} = \mathbf{B} / \sqrt{\mu_0 \rho}$ is the magnetic field in {\alf} units. $\rho = m_p \bar{N_p}$ where $\bar{N_p}$ is an average over a rolling window of 56 seconds of the SWA/PAS density moment.

\begin{equation}\label{eqn: sigmac}
    \sigma_C = \frac{\langle E^+ \rangle - \langle E^- \rangle}{\langle E^+ \rangle + \langle E^- \rangle}
\end{equation} 

$\langle \cdot \cdot \cdot \rangle$ represents a time average over a 60-minute non-overlapping time window, chosen to capture multiple correlation times of magnetic field observations at PSP and Solar Orbiter distances \citep{Parashar-2020, Chen-2020}. $\mathrm{sign} \langle B_{r} \rangle$ is the polarity of the averaged (background) magnetic field.



By separating the data into bins of {\sigmac}, we examine how {$\Sp$} and {$\oxy$} are dependent on the {\alfty} of the wind. Panel (c) shows that wind with {$\oxy$} $\leq$ 0.145, which has been shown as a cutoff between wind emerging from CH (low {$\oxy$}) versus non-CH structures \citep[][]{Zhao-2009},\ is dominated by the highest {\sigmac} values (0.8 $\leq$ {\sigmac} $\leq$ 1.0), consistent with work showing {\alfic} streams emerge from CH and their boundaries. The correlation coefficients demonstrate that the anti-correlation between {$\Sp$} and {$\oxy$} remains consistent across all {\sigmac} bins, indicating that this relationship remains true regardless of the solar wind's {\alfty}.

Notably, the highly {\alfic} wind's distribution is more concentrated, clustering around lower {$\oxy$} and higher {$\Sp$} values. In contrast, the less {\alfic} wind shows a much broader distribution, spanning a wider range of both charge state ratios and specific entropy. This suggests that the less {\alfic} solar wind may originate from more diverse lower coronal source regions, or perhaps undergo more evolution in these parameters over distance. The concentration of highly {\alfic} wind is consistent with a more homogeneous source, such as coronal holes, while the broader distribution of less {\alfic} wind indicates that both of these parameters are potential classifiers of wind type and indicate information about the source region(s) or acceleration mechanisms at work in the various types of solar wind.


\subsection{Dependence on Polytropic Index} 

The definition of specific entropy as defined thus far is based on the assumption of adiabatic evolution of the wind ($\gamma = 5/3$). However, through statistical observations between 0.3 and 1 AU from Helios 1 and 2, it has been established that the solar wind does not simply undergo adiabatic cooling, there is some additional heating as the wind evolves \citep{Marsch-1982, Hellinger-2011}. It has been shown that an iso-poly model, one that combines an isothermal solar wind up to a critical point ($\sim$ 10 {\Rsun}), and then continues to expand through the inner heliosphere (with a non-adiabatic polytropic index) is a better fit to observed solar wind streams on both an individual \citep[][]{Rivera-2024, Rivera-2025SSW}, and statistical basis \citep[][]{Dakeyo-2022}. Therefore, the choice of $\gamma = 5/3$ as used in Figure~\ref{fig: timeseries_correlation} is likely different than the actual $\gamma$ value associated with the expansion of the wind, and we note that this index is both speed and species dependent. $\gamma$ can theoretically be any value, but likely between $1$ (isothermal) and $5/3$ (adiabatic), with statistical work having shown the value ranges typically between 1.35 and 1.60 \citep{Dakeyo-2022}.

\begin{figure}[ht!]
  \includegraphics[width=\columnwidth]{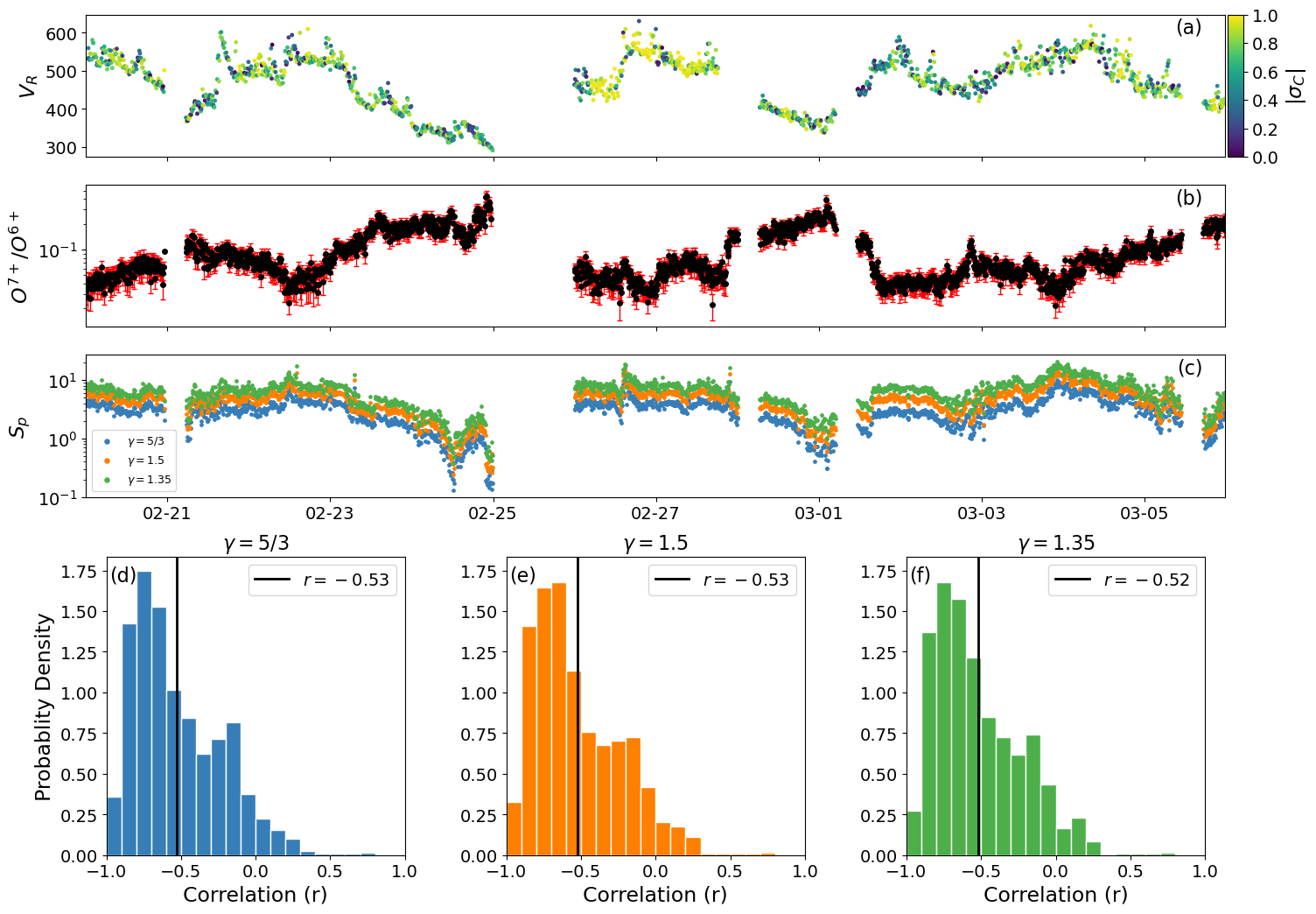}
      \caption{Time series and statistical comparison of the correlation between {$\oxy$} and {$\Sp$} for different polytropic indices. Panel (a) shows a time series of radial solar wind velocity colored by the absolute cross helicity, an indicator of {\alfic} fluctuations. Panel (b) displays the oxygen charge state ratio as measured by SWA/HIS, including measurement uncertainties (shown as red error bars). Panel (c) shows the specific proton entropy determined from SWA/PAS measurements for three different polytropic indices ($\gamma = \frac{5}{3}$ (blue), 1.5 (orange), and 1.35 (green)), illustrating how different assumptions about solar wind thermodynamics influence the relationship between these parameters. Panel (d-f) presents histograms of the rolling correlation between log({$\Sp$}) and log({$\oxy$}) computed over 7-day windows for each from January 2022 through April 2023. The vertical black line in each histogram marks the mean correlation coefficient (also listed in legend).}
  \label{fig: different_gamma}
\end{figure}

To quantify the effect of a polytropic index different than 5/3, we calculated entropy at Solar Orbiter using three different $\gamma$ values, $1.35$ ($\gamma_p$ for the fastest ($v_{sw} \geq 450 \mathrm{km/s}$) solar wind in \citet{Dakeyo-2022}), $1.5$ ($\gamma_p$ for the medium speed ($\sim 350 - 450 \mathrm{km/s}$) wind in \citet{Dakeyo-2022}), and the adiabatic limit $5/3$, to determine if that would affect the entropy correlation with the charge state ratio. In Figure~\ref{fig: different_gamma} we show time series of all three calculations of entropy alongside the {$\oxy$} charge state ratio for a shorter time period of Solar Orbiter observations (February to March 2022). This period has been studied extensively in prior work, and individual streams have been mapped to their sources \citep[e.g.][]{Yardley-2024, Rivera-2024SB, Ervin-2024CH, Rivera-2025SSW}, and is discussed later in the text. We show that one calculation does not align better with the charge state for the entire period. Rather, the change in $\gamma$ rescales the magnitude of the entropy and there maintains a strong relationship between the parameters of interest. The histograms in panels (d) through (f) extend this analysis to the full dataset (January 2022-April 2023), as in Figure~\ref{fig: timeseries_correlation}. This is not surprising as the most appropriate value of $\gamma$ (to describe a specific stream's evolution) likely change across different streams.


In panel (a) we show the velocity timeseries from SWA/PAS colored by the absolute value of the cross helicity. When $|\sigma_C| > 0.7$ we consider the wind to be highly {\alfic}. We see that the faster wind is typically {\alfic} while the slower wind shows more variability. Panel (b) displays the oxygen charge state ratio over the time period and panel (c) shows the corresponding time series of {$\Sp$} for the three different polytropic indices. We see that temporal variations of the parameters are generally similar, suggesting a consistent inverse trend between them over time.

Panels (d) to (f) quantify this relationship through the histograms of the Pearson correlation coefficients between log({$\Sp$}) and log({$\oxy$}), computed over 7-day windows for each $\gamma$ value. The distributions show that the correlation is quasi-consistent regardless of the $\gamma$ value chosen. The mean correlation coefficient is approximately -0.5 for each case, indicating the overall relationship is not affected by the assumed polytropic index. Rather, varying $\gamma$ rescales the magnitude of {$\Sp$} and does not alter the overall relationship with {$\oxy$}. However, while $\gamma$ changes the amplitude of the variations, the most representative values of $\gamma$ are those that hold an approximately constant {$\Sp$} along radial evolution of the wind, similar to the conservation of {$\oxy$} during solar wind expansion.





\section{Correlation of entropy and {$\oxy$} as a function of heliocentric distance} \label{sec:radial}
\begin{figure}[ht!]
    \centering
    \includegraphics[width=\linewidth]{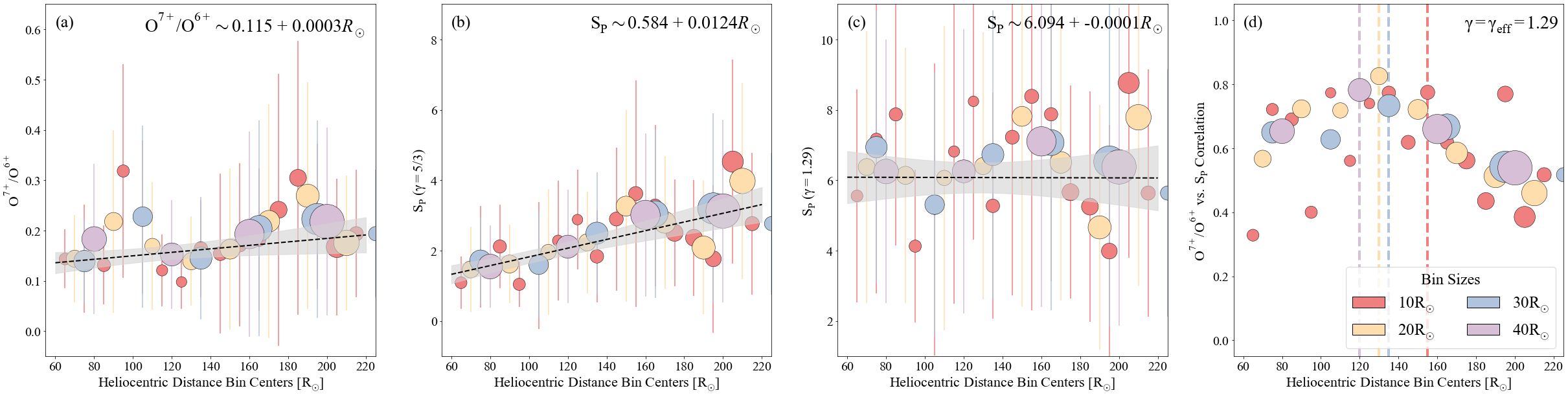}
    \caption{Panel (a) shows the evolution of {$\oxy$} and (b) shows the evolution of {$\Sp$} as a function of distance. A weighted least squares regression is applied to calculate the best fit line (black dashed) and the 95\% confidence interval is shaded in gray. Panel (c) is equivalent but for our $\yeff$ value of 1.29.
    Panel (d) shows the absolute value of the correlation coefficient as a function of distance (plotted against the central distance) for different radial bin sizes. The size of the marker corresponds to the number of samples within each bin (between $\sim$1000 and $\sim$11000) depending on the bin sizes. The dashed vertical lines show the radial location corresponding to the maximum correlation between the parameters. The 10 {\Rsun} bin size also peaks at 135 {\Rsun} (like the 30 {\Rsun} bin size). }
    \label{fig:radial}
\end{figure}

If we are interested in using {$\Sp$} as an in-situ source region proxy, understanding its radial evolution is vital for applying this at near-Sun distances such as those observed by PSP. Solar Orbiter traverses a large span of radial distances during this period, $\sim$0.28 to $\sim$1 AU, ($\sim$60 {\Rsun} to $\sim $ 220 {\Rsun}), providing us with the unique opportunity to study the radial dependence of the {$\Sp$} and {$\oxy$} correlation and gain insight on the utility of Sp as a source proxy at different points in the heliosphere. In Figure~\ref{fig:radial}, we show the evolution of {$\Sp$} and {$\oxy$} over the distances Solar Orbiter measures. We use the four different bin sizes, (10, 20, 30, and 40~{\Rsun}) chosen to ensure that each selected bin had at least 1000 samples. Bin sizes larger than 40 {\Rsun} are unable to capture the evolution of the parameters or the correlation.  

In Figure~\ref{fig:radial}(a) we show the average and standard deviation {$\oxy$} as a function of heliocentric distance for different radial bin sizes. We see that this quantity approximately conserved through radial evolution, as expected. We use a weighted least squares regression to find a line of best fit noting {$\oxy$} $\propto 0.0003 r(R_\odot)$ from the HIS measurements, indicating that this quantity is globally conserved over radial evolution (despite mixing from different wind streams).

In Figure~\ref{fig:radial}(b) we show {$\Sp$} as a function of distance for the different bin sizes. We see that this quantity is not conserved through evolution. The larger radial bins (30 and 40 {\Rsun}) seem to show a saturation of specific entropy at larger distances ($>$ 150 {\Rsun}) that should be investigated in future work.  

We fit this radial trend using a weighted least squares regression to account for the variance in the measurements in each bin. We find that the best fit to the observations has $S_p \sim 0.584 + 0.012 \times r(R_\odot)$. The evolution of {$\Sp$} intrinsically relates to the average non-adiabatic behavior of the wind. From the Parker model, we expect $S_p$ to be conserved as a function of distance, meaning that the growth of $S_p$ as a function of distance (Figure~\ref{fig:radial}(b)) indicates that $\gamma = 5/3$ is not a good description of the wind's evolution \citep{Parker-1958}. The conservation of entropy as a function of distance (provided the correct polytropic index is chosen) also supports the use of $\Sp$ as a charge state proxy.

\begin{align}
    n(r) \sim \frac{1}{r^2} \\
    T(r) \sim n^{\gamma - 1} \\
    S_p(r) \sim \frac{T(r)}{n(r)^{\gamma - 1}} \sim \mathrm{const.}
\end{align}

We determine an ``effective" polytropic index by looking for the $\gamma$ value that gives a constant {$\Sp$} with radial distance over the full datasets of measurements. Using this measurement set, we find that $\gamma = 1.29$ provides the best fit for the conservation of $\Sp$, with a weighted least square fit of $S_p \sim 6.094 - 0.0001 \times r(R_\odot)$, shown in Panel (c). We see that there is a large spread in $\Sp$ values with this $\yeff$, however the radial trend remains relatively constant, especially with larger bin sizes. It is important to note that changing the $\gamma$ value has a significant impact on the magnitude of the $\Sp$ calculated (see Figure~\ref{fig: different_gamma}). Statistical fits from \citet{Rivera-2024, Rivera-2025SSW} to temperature, velocity, and density profiles as a function of distance resulted in $\gamma$ values of 1.4 (fast wind), 1.3 (slow wind) and 1.33 (slow {\alfic} wind). The majority of our measurements are slow and {\alfic} (see Figure~\ref{fig:param-relations}) and thus a value of 1.29 is reasonable and consistent with prior examinations of individual streams, as well as within the range of values from statistical fits to profiles of measurements over varied heliocentric distances \citep{Badman-2025}. We refer to $\gamma = 1.29$ as an ``effective" $\gamma$ value ($\yeff$) moving forward. Using this $\yeff$ value conserves $\Sp$ over radial evolution, similar to the conservation of the $\oxy$ ratio meaning that predictions of $\oxy$ ratios from $\Sp$ values could potentially be used over a large range of heliocentric distances.

In Figure~\ref{fig:radial}(d), we use the same four bin sizes to probe the radial dependence of the {$\Sp$} vs. {$\oxy$} correlation, plotting the absolute value of the correlation coefficient. There are time periods where positive correlation was found ($< 5$\% of the dataset). Similar to the results from \citet{Nakhleh-2025}, we attribute the periods of positive correlation to times when the variability of the solar wind speed was low (e.g. within solar wind streams), with further discussion found in Appendix~\ref{app:correlation}. We find a similar trend in correlation regardless of the bin size used, noting that the correlation peaks between 120 and 140 {\Rsun}. This could potentially be due to the fact that the ``correct" $\gamma$ value to describe the solar wind's radial evolution is not constant as a function of heliocentric distance, or the fact that different solar wind types have different $\Sp$ profiles which could contribute to the non-linear trend in the correlation.
While not the subject of this work, future study looking at why the correlation peaks most strongly at these distances could provide insight as to the thermodynamic properties of the solar wind. 

Multi-point measurements of the evolution of a single plasma parcel would provide constraints as to the appropriate $\gamma$ value for the conservation of entropy in a single stream, while statistical fits could be used to detrend data to predict {$\Sp$} values nearer to the Sun and therefore predict charge state ratios at points without measurement capability. This could also help study what range of $\Sp$ values are associated with different source regions.


\section{Relation to Wind Type and Source}
\label{sec: categorization} 

\begin{figure}[ht!]
    \centering
    \includegraphics[width=\columnwidth]{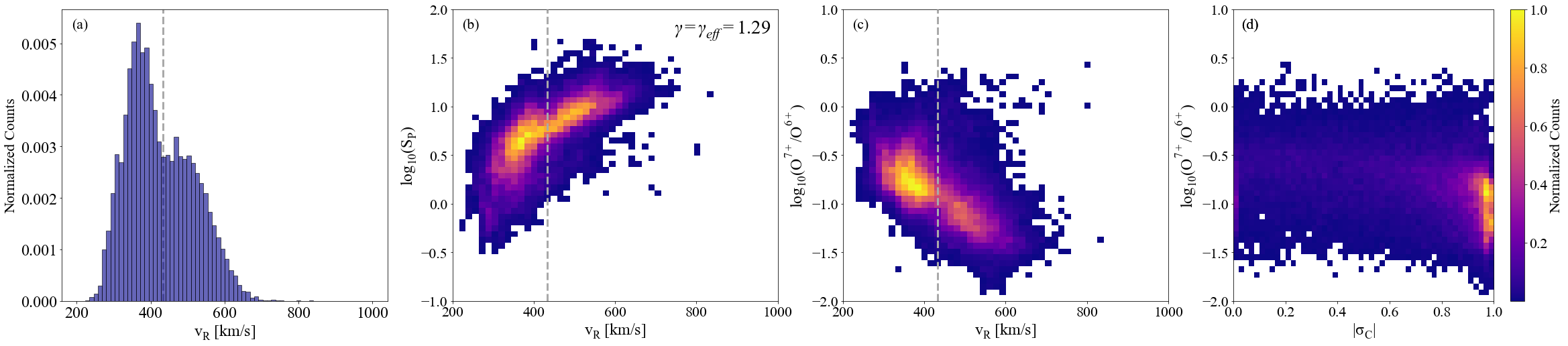}
    \caption{Comparison between in-situ solar wind classification parameters ({\sigmac} and $v_R$) with in-situ metrics of solar wind source regions ({$\Sp$} and {$\oxy$}). Panel (a) shows a 1d distribution of solar wind speeds associated with the time range of interest (Jan 2022 - April 2023). Panel (b) and (c) show normalized 2d distributions of $v_R$ and {$\Sp$} and {$\oxy$}, where $\Sp$ is computed with $\yeff$. The dashed vertical line shows the 423 {\kms} threshold from \citet{Alterman-2025}. Panel (d) shows the 2d distribution of {\sigmac} and {$\oxy$}. Joint distributions are normalized to the maximum count in each distribution.}
    \label{fig:param-relations}
\end{figure}

As our goal is to understand the utility of the specific entropy as a source region classifier, we look to estimate an equivalent threshold for {$\Sp$} to the CH vs. non-CH threshold for the $\oxy$ measurements \citep{Zhao-2009}. As pristine fast solar wind (FSW) is known to emerge from CH structures \citep{McComas-1998, McComas-2008}, we assume that all the (highly {\alfic}) FSW observed by Solar Orbiter emerges from CH structures. We categorize our full dataset based on in-situ characteristics often used for stream categorization: solar wind speed ($v_R$) and cross helicity (\sigmac). In Figure~\ref{fig:param-relations}, we show the relationship between parameters of interest for source region tracing ({$\oxy$} and {$\Sp$}) and parameters we use for solar wind classification ($v_R$ and {\sigmac}), to motivate our classification scheme. 


In Figure~\ref{fig:param-relations}(a), we see that there is a bimodal distribution of wind speeds as seen by SWA/PAS. The vertical dashed line shows the 423 {\kms} \citet{Alterman-2025} threshold for saturation of the alpha-particle abundance. They calculate ``saturation speeds" that should differentiate between source regions (based on the alpha-particle abundance $N_\alpha / N_p$) using various methods. We choose to use their 423 {\kms} value as it corresponds well to the local minimum in the $v_R$ distribution. Panel (b) shows the strong positive relation between $v_R$ and {$\Sp$} and panel (c) shows a negative relation between $v_R$ and {$\oxy$}. We see that panel (c) shows two clusters of $v_R$ to {$\oxy$} correspondence, and that the \citet{Alterman-2025} velocity threshold seems to separate these two groups quite nicely. We note no strong dependence of {$\oxy$} (and thus {$\Sp$}) on {\sigmac} values. 

Based on the distribution of parameters shown in Figure~\ref{fig:param-relations}, we categorize the wind observed by Solar Orbiter (January 2022 through April 2023) into FSW, SASW, and SSW. Solar Orbiter takes measurements between $\sim$ 60 and $\sim$ 210 {\Rsun} during the period of HIS operations and thus the solar wind is expected to have reached asymptotic speeds at these distances. We use 423 {\kms} as a cutoff between fast and slow streams. We also use cross helicity ($|\sigma_C| = 0.7$) to distinguish between {\alfic} and non-{\alfic} wind types. Table~\ref{tab:categorization} describes our wind categorization scheme, and average parameters associated with the wind types.

\begin{table}[ht!]
    \centering
    \begin{tabular}{|c|c|c|c|c|c|}
    \hline
         \textbf{Wind Type} & $\mathbf{V_{SW}}$ & $\mathbf{\sigma_C}$ & $\mathbf{S_P \; (\gamma = 5/3)}$ & $\mathbf{S_P \; (\gamma = \yeff = 1.29)}$ & $\mathbf{O^{7+} / O^{6+}}$ \\
    \hline
    \hline
    \textbf{Fast (FSW)} & $\langle V_{SW} \rangle \geq 423$ km/s & $\langle \sigma_C \rangle \geq 0.7$ & $4.22 \pm 2.24$ & $9.75 \pm 3.94$ & $0.10 \pm 0.12$\\
    \hline
    \textbf{Slow {\alfic} (SASW)} & $\langle V_{SW} \rangle \leq 423$ km/s & $\langle \sigma_C \rangle \geq 0.7$ & $1.58 \pm 1.15$ & $4.65 \pm 2.28$ & $0.20 \pm 0.14$ \\
    \hline
    \textbf{Slow Non-{\alfic} (SSW)} & $\langle V_{SW} \rangle \leq 423$ km/s & $\langle \sigma_C \rangle \leq 0.7$ & $1.28 \pm 1.30$ & $3.52 \pm 2.75 $ & $0.29 \pm 0.17$ \\
    \hline
    \end{tabular}
    \caption{Table outlining the categorization scheme for FSW, SASW, and SSW streams observed by Solar Orbiter. $\langle \cdot \cdot \cdot \rangle$ indicates averaging of the in-situ measurement over a 1-hour period. We report averages and standard deviations for in-situ parameters ({$\Sp$} and {$\oxy$}) over all the wind in the respective category. The $\gamma = \yeff = 1.29$ shows values using $\yeff$ found to be the best-fit at conserving $\Sp$ over distance.}
    \label{tab:categorization}
\end{table}

\begin{figure}
\begin{center}
\includegraphics[width=\columnwidth]{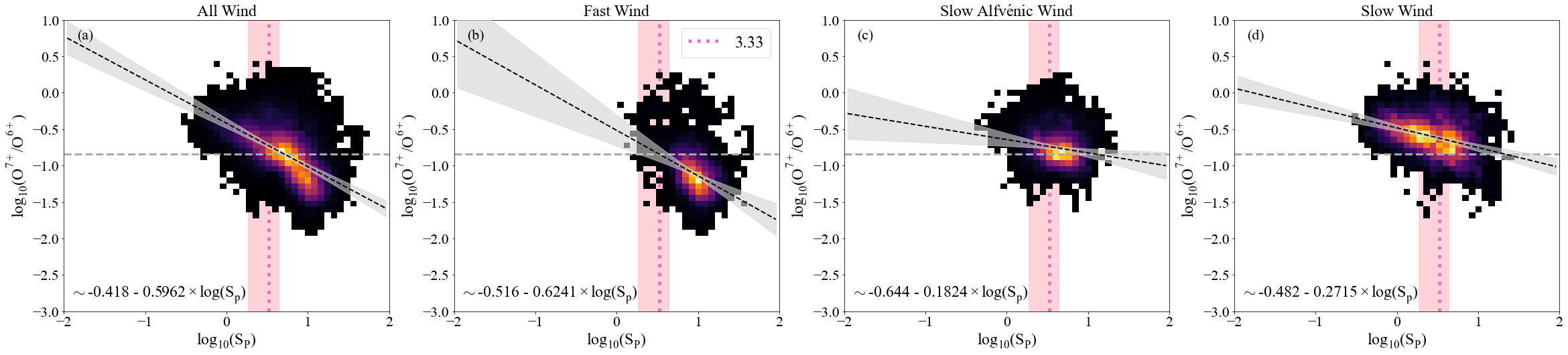}
  \caption{Normalized 2d histograms (to the maximum value in the joint distribution) showing the {$\oxy$} measurement from Solar Orbiter/HIS compared with the {$\Sp$} computed from the MAG and PAS measurements ($\yeff =1.29$). The vertical dashed line at {$\oxy$} = 0.145 is the upper threshold for CH wind \citep{Zhao-2009, Wang-2016}. Panel (a) shows all the wind observed by Solar Orbiter (January 2022 to April 2023). Panels (b), (c), and (d) show the corresponding plot for the fast solar wind, slow {\alfic} wind, and classically slow wind. Wind streams are categorized as in Table~\ref{tab:categorization}. The dashed black lines shows the best fit to the distribution while the gray shading shows the 95\% confidence interval of the least squares fit. The dashed pink lines (pink shaded region) show the intersection of the mean best fit line with the \citet{Zhao-2009} 0.145 threshold for the {$\oxy$} ratio. The shaded pink region corresponds to the overlap with the 95\% (gray shaded) confidence interval of the fit where the vertical dashed pink line is at $\Sp = 3.33$.
  }
  \label{fig: 2dhist_entropy_oxygen}
\end{center}
\end{figure}

In Figure~\ref{fig: 2dhist_entropy_oxygen} we show the correlation between {$\Sp$} and {$\oxy$} for the three different wind types, where $\Sp$ is defined with our radial distance detrended polytropic index ($\yeff = 1.29$). Wind is classified as FSW, SSW, or SASW based on parameters in Table~\ref{tab:categorization}. We calculate fits to the joint distributions using weighted least squares optimization to account for the measurement uncertainty in the {$\oxy$} measurement at each data point. We display the best fit (black dashed line) and a corresponding 95\% confidence interval (gray shading) for each distribution. We experimented with higher order fits, finding no significant difference in the ability of higher order fits to better describe the distribution (see Appendix~\ref{app:correlation}). The horizontal dashed line corresponds to the threshold for the {$\oxy$} charge state ratio to separate wind where a value below 0.145 is CH and above is non-CH origin \citep{Zhao-2009}. 

Based on the best fit line to the joint {$\Sp$} vs. {$\oxy$} distribution, we determine where our best fit line crosses the 0.145 \citet{Zhao-2009} threshold and set this as an ``equivalent" {$\Sp$} threshold. This threshold is shown via the pink dashed line (with uncertainty in pink shading on panel (b) of Figure~\ref{fig: 2dhist_entropy_oxygen}). We find a threshold of $3.33 \pm 1.26$ for $\gamma = 1.29$ ($0.81 \pm 0.47$ for $\gamma=5/3$ using the same analysis).

In panel (a) we show the cross correlation for the full dataset, allowing us to relate {$\oxy$} and {$\Sp$} for the full distribution using $\gamma=1.29$. From our weighted least squares fit, we find the following equation to relate $\Sp$ and $\oxy$ (panel (a)):
\begin{equation} \label{eqn: oxy-sp}
    \log(\oxy) = -0.418 - 0.5962 \times \log(\Sp).
\end{equation}
We use this to predict oxygen charge state values from the Orbiter dataset, finding the average residual to be $0.03 \pm 0.16$ from the measured $\oxy$ value. We primarily see large positive residuals (predicted - measured) of $\oxy$ ratios in fast solar wind. We show and further discuss these results in Appendix~\ref{app:correlation}.  

When looking at the $\Sp$-$\oxy$ relationships between wind types, we find varying results. The FSW (panel (b)) typically has {$\oxy$} ratios that fall below the \lq{}CH-origin\rq{} threshold of \citet{Zhao-2009, Wang-2016} and {$\Sp$} values that fall above the our derived $\Sp$ threshold. This is to be expected as our method for deriving this threshold was based on the assumption that all fast wind was of CH origin. The $\Sp$-$\oxy$ distribution for SASW (panel (c)) lies at the $\oxy$ threshold and above the derived $\Sp$ threshold. This could indicate that the SASW forms a subset or limiting case of the classical fast, {\alfic} wind and emerges from CH boundaries where there may be a mixture of plasma that spent time in nearby hotter, active region structures (leading to larger charge state ratios).

The SSW (panel (d)) shows {$\oxy$} ratios above the \lq{}CH-origin\rq{} threshold indicating that much of this wind likely originates in hot, dense lower corona plasma possibly released via reconnection near the current sheet or the separatrix web \citep{Antiochos-2011}, or through other means. There is a mixture of entropy values associated with the slow wind. As shown in Figure~\ref{fig:param-relations}, the specific entropy ($\gamma = 5/3$) is strongly speed dependent (regardless of polytropic index). Upon inspection of the slow wind with $\Sp > 3.3$, we find that the distribution of speeds here is on the higher end of the slow wind distribution (see Appendix~\ref{app:slow-wind}). This could be due to the fact that some of this wind has not fully evolved to its asymptotic wind speed, and thus our velocity thresholding method identifies wind as ``slow" that may become fast at 1 au. Alternative explanations could relate to measurement uncertainties in the plasma (or heavy ion) data, as well as our choice of polytropic index. Slower winds have been shown to have $\gamma \sim 1.5-1.6$ values \citep{Dakeyo-2022} which could be a possible reason for this.

We use the time period shown in Figure~\ref{fig: different_gamma} to test our {$\Sp$} thresholding approach. \citet{Ervin-2024CH} identified and estimated source region mappings for streams during this time period. They looked at slow wind at the heliospheric current sheet (HCS), a SASW stream that emerged from the edges of a CH, and two fast wind streams emerging from CH structures. In Table~\ref{tab:sp-oxy-e11} we report average {$\oxy$} and {$\Sp$} values for these periods (using $\gamma=1.29$). 

\begin{table}[ht!]
    \centering
    \begin{tabular}{|c|c|c|c|c|c|C|}
    \hline
    \textbf{Wind Stream} & \textbf{Start Time}  & \textbf{End Time} & \textbf{Modeled Source} &
     $\mathbf{{O^{7+} / O^{6+}}}$& 
     $\mathbf{{S_p}}$\\
    \hline
    \hline
    \textbf{FSW (1)}& 2022-03-02 19:12:51 & 2022-03-03 17:09:04 & Equatorial CH& 0.06 $\pm$ 0.01 & 10.17 $\pm$ 2.61\\
   \hline
    \textbf{FSW (2)}& 2022-02-26 14:22:33 & 2022-02-27 12:02:35  & CH Edge/Boundary&  0.05 $\pm 0.02$ & 10.61 $\pm 2.52$\\
    \hline
    \textbf{SASW} &  2022-02-23 00:00:24 & 2022-02-24 23:22:59  & CH Edge/Boundary &  0.17 $\pm 0.06$ & 4.51 $\pm 2.57$\\
    \hline
    \textbf{HCS} &  2022-02-28 19:52:40 & 2022-03-01 04:00:10 &  Current Sheet&   0.25 $\pm 0.03$ & 2.51 $\pm$ 0.68\\   
    \hline
    \end{tabular}
     \caption{Summary of the values associated with the wind streams examined by \citet{Ervin-2024CH}. Two fast solar wind (FSW) streams with high Alfv\'enicity, one slow Alfv\'enic solar wind (SASW) stream, and one stream at the HCS were looked at.}
     \label{tab:sp-oxy-e11}
\end{table}

We demonstrate in Table~\ref{tab:sp-oxy-e11} that the {$\Sp$} of the four different types of streams are statistically different. Both FSW streams and the SASW stream have {$\Sp$} values that fall above our threshold, and beyond the error associated with the {$\Sp$} threshold ($3.33 \pm 1.26$). We note that the SASW stream showed mixed in-situ charge state measurements which could be due to the backmapping methods or a mixing of plasma with a nearby slow wind stream (see \citep{Ervin-2024CH}).

\section{Results and Conclusions}\label{sec: conclusions}

As shown in \citet{Pagel-2004}, with the new Solar Orbiter dataset through comparison of the oxygen charge state ratio ({$\oxy$}) with the specific proton entropy ({$\Sp$}), we show that the specific proton entropy shows a high degree of (anti-)correlation with the oxygen charge state ratio over a wide range of heliocentric distances (0.28 to 1 AU). Therefore, like the charge state measurements, entropy has the potential to be a good tracer of the lower coronal source regions, despite requiring the ``correct" polytropic index to make it a conserved quantity through radial evolution. Regardless, we find that the strength of the relationship between these parameters has little dependence on the value of the polytropic index ($\gamma$) chosen to calculate {$\Sp$} (Figure~\ref{fig: different_gamma}). 

We studied the evolution of the {$\oxy$} and {$\Sp$} correlation as a function of radial distance (from 0.28 to 1 AU) and found that the specific entropy evolution (with $\gamma = 5/3$) can be well constrained with a linear fit ({$\Sp$} $\propto 0.012 \times r({R_\odot})$. Polytropic expansion is expected to conserve {$\Sp$} with distance, meaning that the observation of radial evolution with $\gamma = 5/3$ indicates that non-adiabatic evolution is occurring. We use the fact that $\Sp$ should be conserved (assuming the Parker model) to determine a ``best-fit" $\gamma$ that conserves entropy over distance ($\gamma = 1.29$). Using this value of gamma, we derive a best-fit equation to relate the $\Sp$ and $\oxy$ measurements such that $\Sp$ can be used to predict $\oxy$ when no measurements are available (Equation~\ref{eqn: oxy-sp}). This relationship can be applied to spacecraft without in-situ charge state measurements, and validated during conjunctions with Solar Orbiter at times when HIS was taking observations in future work.

There seems to be a start of entropy saturation around $\sim$150 {\Rsun}, however this depends on the radial bin size used to study the radial evolution. This could also be indicative of where the adiabatic description (e.g. choosing $\gamma = 5/3$) better describes the solar wind evolution. We also find that the correlation between {$\Sp$} and {$\oxy$} peaks at $\sim 130$ {\Rsun}. Further investigation of the evolution of entropy using multi-spacecraft conjunctions could help better constrain this saturation and correlation. 

Using the Solar Orbiter observations, we estimated a threshold for {$\Sp$} to differentiate between CH and non-CH wind. We split the Solar Orbiter observations into fast wind (FSW), slow {\alfic} wind (SASW), and slow non-{\alfic} wind (SSW) following Table~\ref{tab:categorization}. We use a weighted least squares fit to the joint distribution of {$\oxy$} and {$\Sp$} to account for measurement error in {$\oxy$} and find a best fit line to the distribution. We use the overlap between our best fit line and the \citet{Wang-2009} {$\oxy$} threshold separating CH from non-CH wind, as an ``equivalent" {$\Sp$} threshold. We specifically do this for our fast wind category which we canonically assume to emerge only from coronal holes. We find a threshold of $3.33 \pm 1.26$, where the uncertainty accounts for the 95\% confidence interval in our weighted least squares fit. Applying, this thresholds to streams that have been previously studied by \citet{Ervin-2024CH}, we find that it holds in differentiating between CH and non-CH wind (as determined from tracing spacecraft measurements through models of the coronal magnetic field). Further work to reduce the uncertainty on this threshold, and comparing this threshold to other mapped streams (also accounting for different distances) is necessary to determine if it is a viable in-situ proxy for source region.

This study highlights the continued potential for using proton specific entropy as a high cadence proxy for solar wind charge state measurements, at least down to 0.28 AU, particularly in tracing source regions when direct heavy ion charge state measurements are unavailable. Additional work using modeling to connect in-situ observations of the solar wind to their source, and accounting for the varying polytropic index between streams, would help solidify {$\Sp$} as an in-situ proxy for lower coronal source region. Identifying the appropriate polytropic index for conservation of {$\Sp$} through radial evolution is vital to fully apply this proxy at different distances. Using conjunction periods between Parker Solar Probe and Solar Orbiter would allow us to study the validity of using {$\Sp$} as an {$\oxy$} charge state proxy near the Sun, as well as study the radial evolution of specific entropy. This could also be compared with measurements of the electron temperature which has been shown to relate to the {$\oxy$} ratio \citep{Rivera-2024SB}.

\section{Acknowledgments} \label{sec: acknowledgements}
J.D.C. and T.E acknowledge funding from NASA contract NNN06AA01C. T.E. acknowledges funding from The Chuck Lorre Family Foundation Big Bang Theory Graduate Fellowship. Y. J. R and S.T.B. were partially supported by Parker Solar Probe project through the SAO/SWEAP subcontract 975569. N.M.V. is supported by the competed NASA Heliophysics Internal Scientist Funding Model.

Solar Orbiter is a mission of international cooperation between ESA and NASA, operated by ESA. Funding for SwRI was provided by NASA contract NNG10EK25C. Funding for the University of Michigan was provided through SwRI subcontract A99201MO.

This research used version 4.1.6 of the SunPy open source software package \citep{sunpy}, and made use of HelioPy, a community-developed Python package for space physics \citep{heliopy}.

\software{
\texttt{Astropy} \citep{astropy:2013, astropy:2018, astropy:2022},
\texttt{heliopy} \citep{heliopy},
\texttt{matplotlib} \citep{mpl},
\texttt{numpy} \citep{numpy},
\texttt{pandas} \citep{pandas},
\texttt{pySPEDAS} \citep{SPEDAS},
\texttt{scipy} \citep{scipy},
\texttt{spiceypy} \citep{spiceypy},
\texttt{SunPy} \citep{sunpy}
}

\bibliography{ms}{}
\bibliographystyle{aasjournal}

\appendix
\section{Solar Orbiter Data Cleaning Process} \label{app:data-cleaning}
Solar Orbiter measurements from the fluxgate magnetometer (MAG), proton and alpha sensor (PAS), and heavy ion sensor (HIS) were used in this work. These instruments produce measurements at different cadences, and a merged dataset was necessary to be produced for this study. We outline our data processing pipeline for producing this merged dataset below.

The raw data used from each instrument is as follows:
\begin{enumerate}
    \item \textbf{MAG}: raw L2 data at 1 second cadence from which we access $\mathrm{B_{RTN}}$. We use the instrument quality flag to mask out bad data points. 
    \item \textbf{PAS}: raw L2 data at 4 second cadence from which we access $\mathrm{v_{RTN}}$ and $\mathrm{N_P}$. We use the instrument quality flag to mask out bad data points. 
    \subitem (a) {$\Sp$} is calculated from this cleaned dataset. 
    \item \textbf{HIS}: raw L3 data at a 10 minute cadence from which we access the {$\oxy$} ratio and associated measurement error. We use the instrument quality flag to mask out bad data points, choosing to only keep the \texttt{quality\_flag=0} data. 
\end{enumerate}

Two merged datasets were then produced:
\begin{enumerate}
    \item \textbf{Merged Plasma Dataset (PAS/MAG)}: MAG and PAS data are merged to the PAS (lower cadence) time stamps. We take the average over MAG observations between each PAS measurement, in non-overlapping windows. 
    \subitem (a) This merged dataset is used to calculate {\sigmac}, where we used an average of the $N_P$ measurement. $\bar{N_p}$ is an average over a rolling window of 56 seconds (16 observations) of the cleaned SWA/PAS proton density measurement. This was calculated as an additional parameter in the Merged Plasma Dataset.
    \item \textbf{Full Merged Dataset (HIS/PAS/MAG)}: The merged plasma dataset was downsampled to the HIS (lower cadence) time stamps. We take a non-overlapping average over the merged observations between each HIS measurement. 
    \subitem (a) This merged dataset is used for investigation of the {$\oxy$} and {$\Sp$} relation.
\end{enumerate}

This full merged dataset included observations from Jan 2022 through March 2023. After data cleaning and downsampling of measurements to the HIS cadence, we had 30402 observations of {$\Sp$} and {$\oxy$} over this time period. We note a large data gap (see Figure~\ref{fig: timeseries_correlation}(a)) spanning from June 23 to August 26, 2022 due to the quality flag associated with the HIS measurements. 

\section{Prediction of $\oxy$ from $\Sp$ measurements} \label{app:correlation}

Using our expression to relate specific entropy and charge state ratios (Equation~\ref{eqn: oxy-sp}), we predict {$\oxy$} values over the full dataset from our measurements of $\Sp$. We use both the linear fit (shown in the main text) and a second order fit of the joint distribution to do the fitting. We find no significant difference in our ability to predict the $\oxy$ charge state ratio  from the second order fits.

\begin{figure}
\begin{center}
\includegraphics[width=\columnwidth]{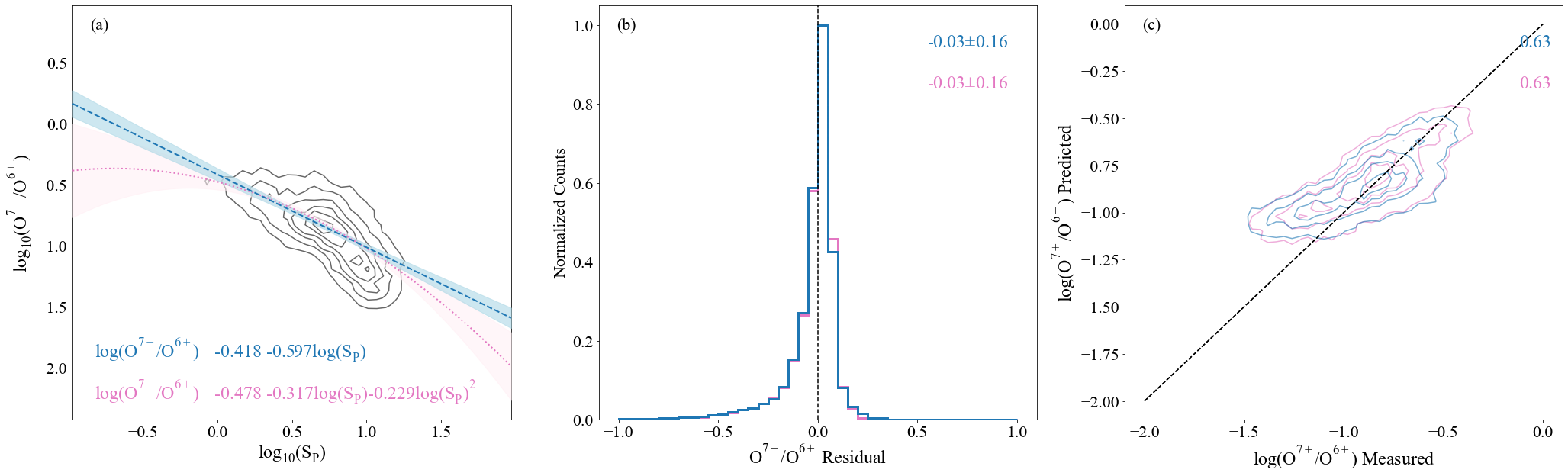}
  \caption{Comparison of ``predicted" and measured $\oxy$ values using Equation~\ref{eqn: oxy-sp}. Panel (a): joint distribution of $\Sp$-$\oxy$ and associated first (blue) and second (pink) order weighted least square fits. Panel (b): distribution of residual values when using the first and second order fits to predict $\oxy$ from $\Sp$. Panel (c): joint distribution of the ``predicted" and measured $\oxy$ ratio. The dashed line shows a $y = x$ line.
  }
  \label{fig: slow_entropy}
\end{center}
\end{figure}

We note that there is a population of wind where we are over-predicting the $\oxy$ value compared to what is measured (bottom left of the joint distribution). We find that this is primarily fast wind, where $v_R \sim 504 \pm 71$ \kms. It is plausible that this is due to an instrumental effect at these higher speeds. \citet{Ogilvie-1985} found a similar effect in ICI measurements of heavy ions in the solar wind where $\mathrm{O^{7+}}$ measurements at higher speeds were contaminated by $\mathrm{He^{++}}$
ions leading to large $\oxy$ ratios than would be expected from coronal source region temperatures. Further investigation of this is necessary to understand the if this instrumental, or if another way to describe the $\oxy$-$\Sp$ relationship is necessary for higher velocity streams. 

\begin{figure}
\begin{center}
\includegraphics[width=\columnwidth]{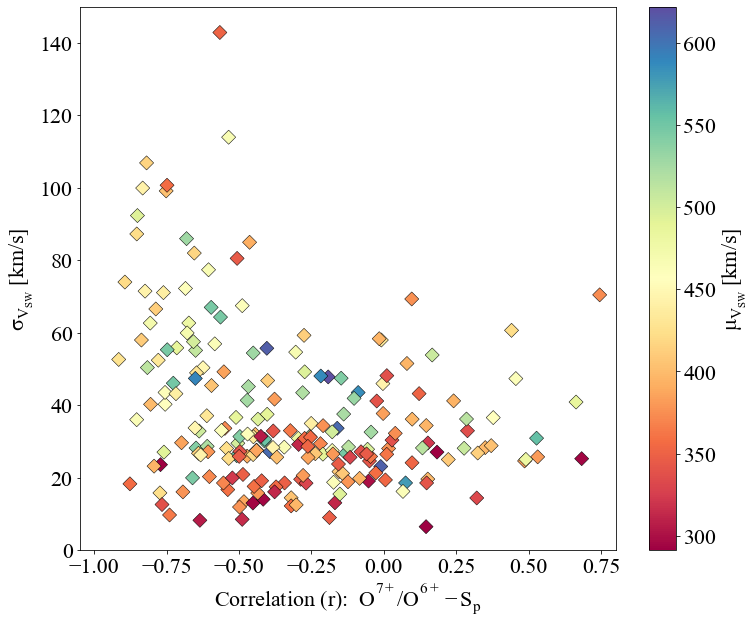}
  \caption{Comparison of the $\Sp$-$\oxy$ correlation coefficient over a 1-day period with the standard deviation ($\sigma$) of the solar wind speed in that same period. The markers are colored by the median solar wind speed in that 1-day interval.
  }
  \label{fig: sigma_vsw}
\end{center}
\end{figure}

Similar to \citet{Nakhleh-2025}, we find that the correlation (and thus the prediction) is stronger when the standard deviation of the wind speed is larger (Figure~\ref{fig: sigma_vsw}). This indicates that $\Sp$ does a better job at tracking changes in $\oxy$ between streams / sources, rather than tracking $\oxy$ within a single stream. 

\section{Discussion of high $\Sp$ slow wind} \label{app:slow-wind}
In Figure~\ref{fig: 2dhist_entropy_oxygen}, we see that some of the slow wind shows entropy that is higher than expected based on the $\oxy$ measurement. We find that the majority of this ``high-entropy" slow wind is on the faster end of all the slow wind observed (see Figure~\ref{fig: slow_entropy}). Figure~\ref{fig:param-relations} shows that entropy is speed dependent and this is consistent with that. 

\begin{figure}
\begin{center}
\includegraphics[width=\columnwidth]{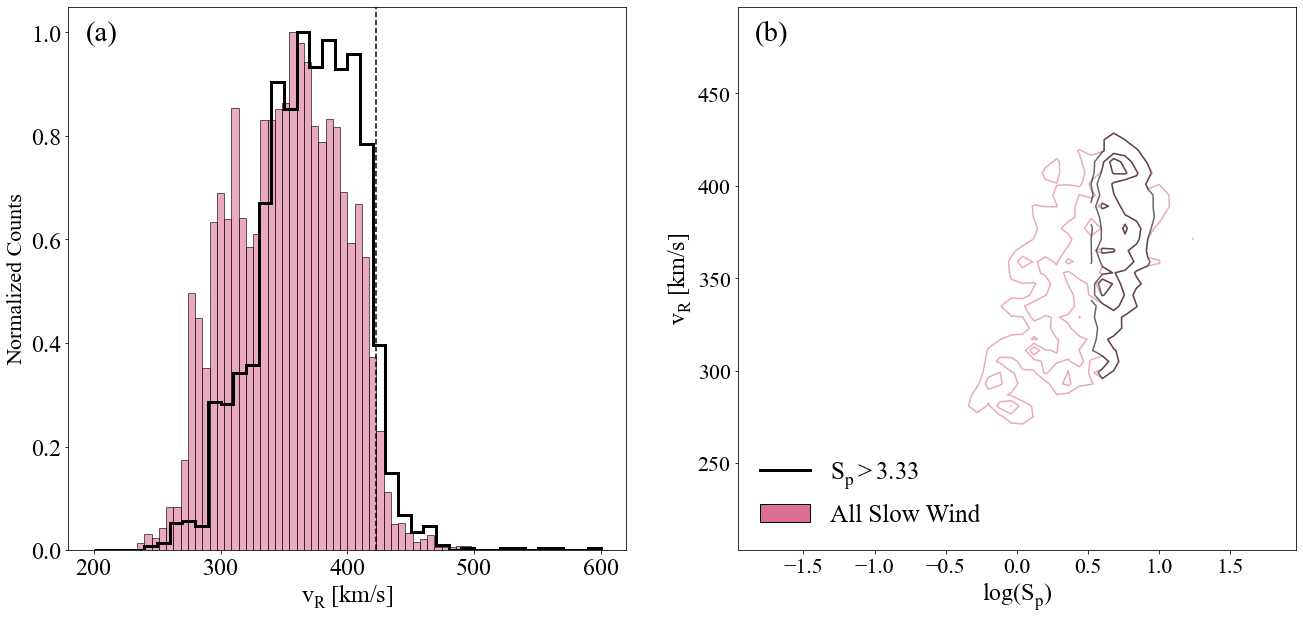}
  \caption{Panel (a): 1d distribution of speeds for all wind categorized as non-{\alfic} slow wind (pink) and for slow wind with $\Sp > 3.33$ (black). Panel (b): 0.25, 0.50, and 0.75 contours of 2d joint distributions of $\mathrm{\log(\Sp)}$ versus $v_R$ for all (pink) and high entropy (black) slow wind.
  }
  \label{fig: slow_entropy}
\end{center}
\end{figure}

\end{document}